\documentstyle [12pt,eqsecnum,aps,amsfonts] {revtex}
\input epsf
\newcommand{\beq}{\begin{equation}}
\newcommand{\eeq}{\end{equation}}
\newcommand{\beqs}{\begin{eqnarray}}
\newcommand{\eeqs}{\end{eqnarray}}

\begin{document}

\draft

\baselineskip 6.0mm

\title{Complex-Temperature Partition Function Zeros of the Potts Model 
on the Honeycomb and Kagom\'e Lattices}

\vspace{4mm}

\author{Heiko Feldmann\thanks{email: feldmann@insti.physics.sunysb.edu}, 
Robert Shrock\thanks{email: shrock@insti.physics.sunysb.edu} 
\and Shan-Ho Tsai\thanks{email: tsai@insti.physics.sunysb.edu}}

\address{
Institute for Theoretical Physics  \\
State University of New York       \\
Stony Brook, N. Y. 11794-3840}

\maketitle

\vspace{4mm}

\begin{abstract}
  
    We calculate complex-temperature (CT) zeros of the partition function for 
the $q$-state Potts model on the honeycomb and kagom\'e lattices for several
values of $q$.  These give information on the CT phase diagrams.  A comparison
of results obtained for different boundary conditions and a discussion of some
CT singularities are given.  Among other results, our findings show that 
the Potts antiferromagnet with $q=4$ and $q=5$ on the kagom\'e lattice has no 
phase transition at either finite or zero temperature. 

\end{abstract}


\vspace{10mm}

\pagestyle{empty}
\newpage

\pagestyle{plain}
\pagenumbering{arabic}
\renewcommand{\thefootnote}{\arabic{footnote}}
\setcounter{footnote}{0}

\section{Introduction}

  The 2D $q$--state Potts models \cite{potts,wurev} for various $q$ have been 
of interest as examples of different universality classes for phase transitions
and, for $q=3,4$, as models for the adsorption of gases on certain substrates.
The $q=2$ Ising special case has long served as a simple exactly solvable 
model of cooperative phenomena.  However, for $q \ge 3$, the free energy of the
Potts model has never been calculated in closed form for arbitrary 
temperature.  It is thus worthwhile to obtain further information about the
properties of the Potts model, and we shall do this in the present paper via 
calculations of complex-temperature (CT) zeros of the partition function of 
the Potts model for the honeycomb and kagom\'e lattices.  
One of the motivations for this work is the recent calculation and 
analysis of long low-temperature series for the $q$-state Potts model on 
these lattices by Jensen, Guttmann, and Enting \cite{jge}.  Our results enable
one to relate the CT singularities in thermodynamic quantities found in Ref. 
\cite{jge} to positions on the CT phase boundaries of the respective models. 

   The study of statistical mechanical models with magnetic field \cite{yl},
temperature \cite{mef}-\cite{dg}, or both \cite{ih} generalized from
real to complex values has yielded interesting insights into the properties
of these models. For a discrete spin model at temperature $T$ and in an
external magnetic field $H$, the partition function $Z$ is, up to a 
prefactor, a 
polynomial in the Boltzmann weights $z(K)$ and $\mu(h)$ containing 
dependence on $K=\beta J$ and $h=\beta H$, where $\beta = 1/(k_BT)$, and $J$ 
is the spin-spin coupling.  It is of interest to study the zeros of $Z$ (i) 
in the complex $\mu$ plane for physical $T$ \cite{yl}; (ii) in the complex $z$
plane for physical (vanishing or nonvanishing) $H$ \cite{mef}; and (iii) on the
${\Bbb C}^2$ manifold $(\mu,z)$ when both $K$ and $h$ are complex \cite{ih}. 
Here we shall concentrate on case (ii), i.e., Fisher zeros.  In the 
thermodynamic limit, via a coalescence of zeros, there forms a continuous 
locus ${\cal B}$ of points where the free energy is nonanalytic. 
This locus serves as the union of boundaries 
(whence the symbol ${\cal B}$) of the various complex-temperature phases. 
Thus, calculations of complex-temperature partition function zeros on 
sufficiently large finite lattices yield useful information on the CT phase 
diagram in the thermodynamic limit.  (Hereafter, to avoid repetition, we shall
simply refer to zeros of the partition function, it being understood that these
are complex-temperature zeros.)  In making
inferences from such finite-lattice calculations about ${\cal B}$ in the
thermodynamic limit it is important to vary both the lattice size and the type
of boundary conditions to have an accurate idea of the sensitivity of the
locations of the zeros to these choices.  Some of the earliest work on CT
properties of spin models dealt with these zeros \cite{mef,kat}.  Another
major reason for early interest in these properties of spin models was the 
fact that unphysical, CT singularities complicated the analysis of
low-temperature series expansions to get information about the location
and exponents of the physical phase transition \cite{dg}.  A third reason for 
interest in these properties is the fact that, as additional sources of
information about thermodynamic functions, they can expedite progress toward
exact solutions.  Aside from well-understood exceptions \cite{com}, 
CT singularities of thermodynamic functions occur on the continuous locus of 
points ${\cal B}$ where the free energy is nonanalytic.  Hence, when 
investigating these singularities, it is useful
to do so in conjunction with a calculation of the zeros of the partition
function to infer the approximate location of the phase boundary 
${\cal B}$ separating various CT phases \cite{cte}.  Interestingly, 
some of these singularities can be related directly to physical
singularities: by using duality, one can show an exact equivalence of the free 
energy of the $q$-state Potts antiferromagnet on a lattice $\Lambda$ for the 
full temperature interval $0 \le T \le \infty$ and the free energy of the 
$q$-state Potts model on the dual lattice for a semi-infinite interval of 
complex temperatures \cite{hcl}. This implies the existence of two quite 
different types of CT singularities: the generic kind, which does not obey 
universality or various scaling relations \cite{chisq,ms}, and a special 
kind which does obey such properties
and encodes information of direct physical relevance.  Although we
consider the honeycomb and kagom\'e lattices here, we mention that previous
calculations of zeros of the partition function for the Potts model with 
$q \ge 3$ have been done on the triangular and square lattices 
\cite{mm,martinbook,wuz,pfef}.  

\section{Model}

The (isotropic, nearest-neighbor) $q$-state Potts model on a lattice $\Lambda$
is defined by the partition function
\beq
Z = \sum_{ \{ \sigma_n \} } e^{-\beta {\cal H}}
\label{zfun}
\eeq
with the Hamiltonian
\beq
{\cal H} = -J \sum_{\langle nn' \rangle} \delta_{\sigma_n \sigma_{n'}}
- H \sum_n \delta_{1 \ \sigma_n }
\label{ham}
\eeq
where $\sigma_n=1,...,q$ are ${\Bbb Z}_q$-valued variables on each site 
$n \in \Lambda$, $\beta = (k_BT)^{-1}$, and $\langle n n' \rangle$
denotes pairs of nearest-neighbor sites.  The symmetry group of the Potts
Hamiltonian is the symmetric group on $q$ objects, $S_q$. 
We use the notation introduced above, $K = \beta J$,  $h= \beta H$, and 
\beq
a = z^{-1} = e^{K}
\label{a}
\eeq
\beq
x = \frac{e^K-1}{\sqrt{q}}
\label{x}
\eeq
The (reduced) free energy per site is denoted as 
$f = -\beta F = \lim_{N_s \to \infty} N_s^{-1} \ln Z$, where
$N_s$ denotes the number of sites in the lattice.
There are actually $q$ types of
external fields which one may define, favoring the respective values
$\sigma_n=1,..,q$; it suffices for our purposes to include only one.
The order parameter (magnetization) is defined to be $m = (qM-1)/(q-1)$. 
where $M = \langle \sigma \rangle = \lim_{h \to 0} \partial f/\partial h$.  
With this definition, $m=0$ in the
symmetric, disordered phase, and $m=1$ in the limit of saturated
ferromagnetic (FM) long-range order. We consider the zero-field model, $H=0$. 
For $J >0$ and the dimensionality of interest here, $d=2$, 
the $q$-state Potts model 
has a phase transition from the symmetric, high-temperature paramagnetic (PM) 
phase to a low-temperature phase involving spontaneous breaking of the 
$S_q$ symmetry and onset of ferromagnetic (FM) long-range order.  This
transition is continuous for $2 \le q \le 4$ and first order for $q \ge 5$.  
The critical exponents and universality classes of the cases where
the model has second-order transitions are well understood \cite{wurev,cft}. 
The $q$-state Potts model has the property of duality 
\cite{potts,wurev,kihara,kj}, 
which relates the partition function on a lattice $\Lambda$ with
temperature parameter $a$ to the partition function on the dual lattice with 
temperature parameter 
\beq
a_d \equiv {\cal D}(a) = \frac{a+q-1}{a-1} \ , \quad i.e.\quad
x_d = \frac{1}{x}
\label{ad}
\eeq
Other 
exact results include formulas for the PM-FM transition temperature 
on the square, triangular, and honeycomb lattices \cite{potts,wurev,kj}, and 
calculations of the free energy at the phase transition temperature, and of the
related latent heat for $q \ge 5$ \cite{baxterf}. No formula is known for the
PM-FM transition on the kagom\'e lattice, although there have been a number of
conjectures; for a recent discussion, see Ref. \cite{jge}.  
The case $J < 0$, i.e., the Potts antiferromagnet (AF) has also been of
interest because of its connection with graph colorings and the fact that,
for certain lattices and values of $q$, it exhibits nonzero ground state
entropy \cite{lieb,baxter87}; for a recent discussion, see \cite{ww,wn} and 
references therein. 
Depending on the type of lattice and the value of $q$, the model 
can also have a phase with AFM long-range order.  For $q \ge 3$ 
on the honeycomb lattice there is no AFM phase \cite{sokal,p3afhc}. 
For any lattice $\Lambda$, the partition function can be expressed
in a form involving a sum of powers of $q$ which allows a generalization from
positive integer $q$ to real (or, indeed, complex) $q$, and we shall use the 
generalization to real $q$ at certain places below. 
Reviews of the model include Refs. \cite{wurev,martinbook}. 

   On a finite lattice, the $q$-state Potts model partition function $Z$ is a
polynomial in the Boltzmann weight $a$.  We calculate this polynomial by
transfer matrix methods.  
This is a challenging numerical problem for large lattices, since the degree of
the polynomial is equal to the number of bonds, $N_b = (\Delta/2)N_s$, where
$\Delta$ is the coordination number, and there is a very large range in the
sizes of the coefficients, from $q$ for the highest-degree term $a^{N_b}$ to
exponentially large values for intermediate terms.  The latter property is
obvious from the fact that for $K=0$, i.e., $a=1$, the sum of the 
coefficients in $Z$ is $q^{N_s}$.  From this, we then compute the zeros.  
A general property of the CT phase boundary for any lattice and 
$q$ value is invariance under complex-conjugation:
${\cal B} \to {\cal B}$ as $a \to a^*$.  

   In addition to the locations of the curves comprising the CT phase boundary
${\cal B}$ inferred in the thermodynamic limit from the zeros calculated 
on finite lattices, one can extract further information.  As one approaches the
thermodynamic limit, so that one can define a density of zeros, this density 
normally behaves near a singular point $a_s$ as \cite{mef,abe}
\beq
g(s) \sim s^{1-\alpha} \ , \quad {\rm as} \quad s \to 0
\label{density}
\eeq
where $s$ denotes the arclength along ${\cal B}$ away from $a_s$ (so that
$s=|a-a_s|$ as $s \to 0$) and where the singularity in the free energy at 
$a_s$ is $f_{sing} \sim |a-a_s|^{2-\alpha}$ \cite{aap}.  If the partition
function has a zero at some point $z_0$ with a multiplicity proportional to 
the number of lattice sites, $N_s$, then this formula, eq. (\ref{density}), 
is modified by the
addition of a term proportional to a delta function $\delta(s)$.  In
Ref. \cite{cmo} it was proved (as Theorem 6) that for the Ising model on a 
lattice with odd coordination number, this happens at $z=-1$.  In particular,
this occurs for the Ising model on the honeycomb lattice (see further below). 

\section{Partition Function Zeros on the Honeycomb Lattice}

\subsection{Comparison with Exact ${\cal B}$ for Ising Case}

   In order to study the effects of the finite lattice size and of different 
boundary conditions, as well as checking the computer programs used, 
it is valuable to calculate the zeros for the $q=2$ Ising case
where the resulting locus of zeros can be compared with the exactly known CT
phase boundary ${\cal B}$.  As noted above, these zeros, like the others to 
be presented further below, are calculated by a transfer matrix method.
 From the known expression for the free energy, this
boundary was determined in Ref. \cite{chitri}; it is the locus of solutions to
the equation
\beq
1-2a+6a^2-2a^3+a^4- 2a(1-a)^2 p =0
\label{hcisingboundary}
\eeq
where $-3/2 \le p \le 3$ \cite{pnote}.  
This locus is shown in Figs. \ref{hex2} and 
\ref{hex2special}.  Because $q=2$ and
the honeycomb lattice is bipartite, the CT phase boundary ${\cal B}$ and also 
the set of zeros are invariant under the inversion map\cite{anote}. The CT 
boundary consists of the union of two parts.  The first is 
an arc of the unit circle extending from $\theta=\arg(a)=\pi/3$
around through $a=-1$ to $\theta=-\pi/3$, while the second is 
a lima bean-shaped curve that crosses the
positive real axis at the PM-FM critical point, 
$a_{PM-FM,q=2}=2+\sqrt{3}=3.732...$ and at the PM-AFM critical point, 
$a_{PM-AFM,q=2}=a_{PM-FM,q=2}^{-1}=2-\sqrt{3}=0.267949...$.  These two parts
intersect each other at multiple points at $\pm i$; these multiple points are 
singular points of ${\cal B}$ in the sense of algebraic geometry \cite{alg}. 
The phase surrounding the origin in
the $a$ plane is the AFM phase; the one surrounding the infinite-temperature
point $a=1$ is the PM phase, and the one extending outside ${\cal B}$ to
complex infinity in all directions is the FM phase \cite{cte}.  

\begin{figure}
\centering
\
\epsfxsize=8.5cm
\epsfbox{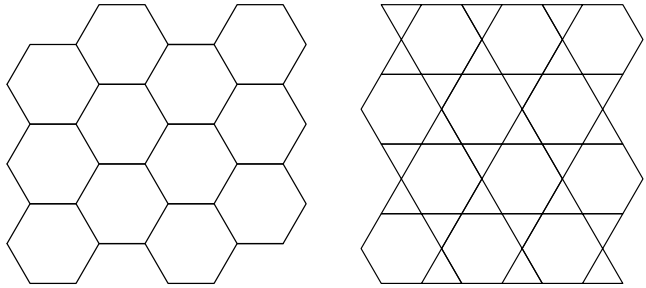}
\caption{Honeycomb and kagom\'e lattices to illustrate our 
conventions for indicating sizes.}
\label{lattices}
\end{figure}

\begin{figure}
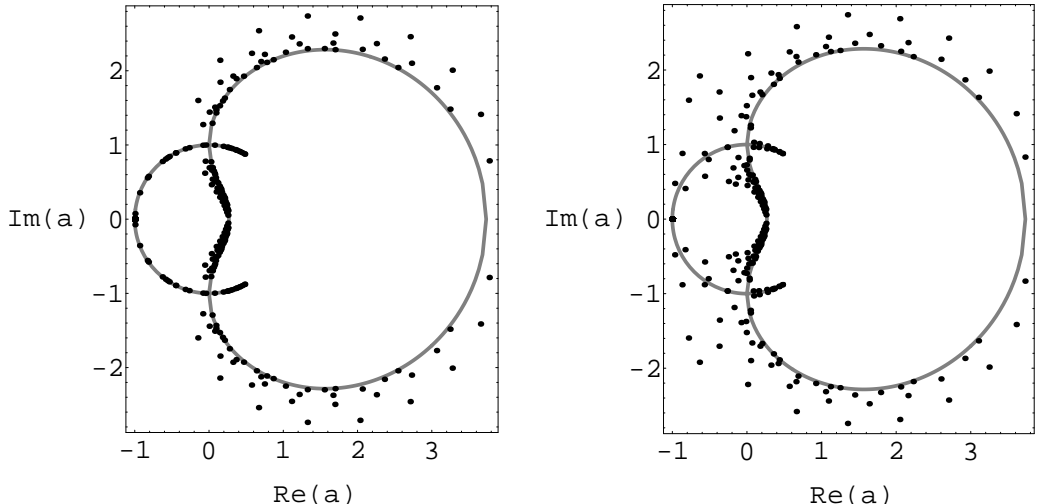

\centering
\
\epsfxsize=6.5 cm
\epsfbox{hex2fp.epsi}
\hspace{0.5 cm}
\epsfxsize=6.5 cm
\epsfbox{hex2pf.epsi}
\caption{Partition function zeros in the complex $a$ plane for the Ising 
($q=2$ Potts) model on a honeycomb lattice. 
(a) left: size $9 \times 12$ hexagons and (fbc, pbc); 
(b) right: $10 \times 10$ hexagons and (pbc, fbc).}
\label{hex2}
\end{figure}

\begin{figure}
\centering
\
\epsfxsize=6.5 cm
\epsfbox{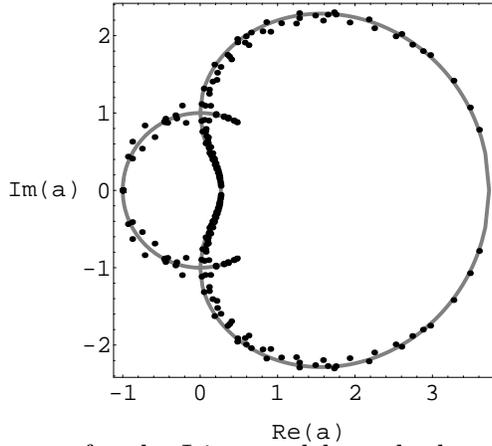}
\caption{Partition function zeros for the Ising model on the honeycomb 
lattice of size $10 \times 10$ hexagons and (pbc, fbc)$_\Delta$ BC's.}
\label{hex2special}
\end{figure}

 Before we start to present our results, we have to introduce our notation for 
the sizes and orientations of the lattices.  We recall first that the 
(infinite) honeycomb lattice is a homopolygonal member, and the kagom\'e 
lattice, a heteropolygonal, member of the class of Archimedean lattices, 
i.e., regular tiling of the plane by one or more types of regular polygons 
such that every vertex is equivalent to every other vertex \cite{gs}.  
An Archimedean lattice is thus uniquely defined by the ordered sequence of 
polygons that one traverses in making a circuit of any vertex.  
In standard mathematical notation 
\cite{gs}, such a lattice is denoted $\Lambda = (\prod_i p_i^{a_i}$) where 
$p_i$ refers to the type of polygon and $a_i$ denotes the number of times 
that it appears consecutively in the product.  In this notation, the 
kagom\'e lattice is denoted $(3 \cdot 6 \cdot 3 \cdot 6)$.  The homopolygonal
subset of Archimedean lattices (consisting of tilings with only one type of
regular polygon) is closed under duality, but the heteropolygonal Archimedean 
lattices (consisting of regular tilings using more than one type of polygon)
have duals that are not Archimedean lattices.  In particular, the dual of the
kagom\'e lattice, called the diced lattice, is not Archimedean.  This
mathematical background will be useful below when we give results for the diced
lattice.   To indicate the size of a given 
lattice for both the honeycomb and kagom\'e cases, we count the number of 
hexagons.   As an illustration, the sizes of the honeycomb and kagom\'e
lattices in Fig. \ref{lattices} are $4 \times 3$ and $3 \times 4$ hexagons,
respectively.  The number of sites in a lattice is also dependent on the 
boundary conditions: with 
periodic boundary conditions in the horizontal direction for example,
the sites on the left and right are identified, while 
with free boundary conditions they
are counted independently from each other.  As is evident from Fig. 
\ref{lattices}, a honeycomb lattice of size, in our notation, $N_x \times N_y$,
is maximally square-like if one takes $N_x$ slightly larger than $N_y$. 

Since we use duality at many points in this work, we chose lattices that have 
natural dual lattices. This excludes 
lattices that are periodic in both directions, for the following reason:
duality relies on the fact that every closed polygon divides the 
lattice into at least two regions.  However, a lattice with periodic boundary
conditions in both directions, and hence with toroidal geometry, has the
property that there exist closed contours that do not divide the surface into 
two disjunct regions.  Since boundary effects are, in 
general, best suppressed if one uses periodic boundary conditions in as many
directions as possible, we use boundary conditions that are periodic in one 
direction and free in the other, i.e., cylindrical boundary conditions. 
Our notation for the boundary conditions
(BC's) is (fbc,pbc) for free and periodic BC's in the horizontal ($x$) and
vertical ($y$) directions, respectively (see Fig. \ref{lattices}), and 
(pbc,fbc) for periodic and free BC's in the $x$ and $y$ directions.  This
notation makes explicit the direction in which the cylindrical boundary
conditions are periodic.

   Complex-temperature partition function zeros for the $q=2$ Ising case of the
Potts model on the honeycomb lattice are shown in Fig. \ref{hex2} for both 
(fbc,pbc) and (pbc, fbc).  The gray curves are the exactly known CT phase
boundary ${\cal B}$.  We remark on several features.
\begin{itemize}

\item The partition function has a multiple zero at $z=-1$ with multiplicity
$\propto N_s$ for large $N_s$ 
This follows from Theorem 6 of Ref. \cite{cmo} and corresponds to the
term $\propto (1-z^2)^{-2}$ in the expression for the specific heat $C$ in the
FM phase given as eq. (3.12) in Ref. \cite{chitri}; the apparent additional
singularity at the infinite-temperature point $z=1$ is not relevant since 
the formula does not apply in that region.  

\item The zeros lie very close to the arcs protruding into the PM phase. 

\item There seems to be some repulsion of the zeros from the multiple points at
$a=\pm i$ (similar to what was seen in Ref. \cite{ih}). 

\item For the (fbc, pbc) case, the zeros on the unit circle show no radial 
deviation.

\item In general, the zeros calculated with the choice (fbc, pbc) lie closer to
the exact boundary curves ${\cal B}$ than those calculated with (pbc, fbc).

\item The zeros lie on the outer side of the boundary between the PM and FM 
phase.  This can be understood as a consequence of the fact that with either
the (fbc,pbc) or (pbc,fbc) boundary conditions, the sites on the free boundary
have a coordination number of two rather than the usual $\Delta=3$ for sites on
an infinite honeycomb lattice.  Hence, ordering effect of the spin-spin 
interactions is commensurately reduced, thereby reducing the finite-lattice
manifestations of the ordered, FM phase, i.e. shifting the PM-FM boundary
outwards. 
\end{itemize}
Note that for the (fbc, pbc) choice, there is one site per hexagon at the 
boundary with only two instead of the usual $\Delta=3$ bonds. 
For the (pbc, fbc) BC's, there are two of these sites for each of the 
hexagons on the upper and lower boundaries. This motivated us to formulate a 
third kind of boundary condition: starting from the (pbc, fbc) choice, 
we added bonds connecting the boundary sites
with fewer than three bounds so that all sites on the lattice have the same 
coordination number $\Delta = 3$.  We denote this choice as
(pbc,fbc)$_\Delta$.  The zeros calculated with this third
choice of boundary conditions are plotted in Fig. \ref{hex2special}.  The main
difference relative to the previous two choices of BC's is that the zeros in
the $Re(a) \ge 0$ half plane have less scatter, lie closer to the exact
boundary ${\cal B}$, and also, in some cases, lie inside the PM-FM phase
boundary.  The zeros with $Re(a) \le 0$ are less scattered than
those with the (pbc,fbc) choice and track the arc of the unit circle well, 
although they do not, in general, lie on it, as was the case with the choice
(fbc,pbc).  The conclusion from this comparison with exactly known results is
that, if one did not know the exact boundary ${\cal B}$ to begin with, one 
would be able, by combining results on zeros calculated with different 
boundary conditions, to reconstruct
it with reasonable accuracy.  

   The density of zeros on ${\cal B}$ near 
the physical PM-FM transition is consistent with vanishing according to
eq. (\ref{density}) with $\alpha=0$, i.e., $g \sim |a-a_{PM-FM,q=2}|$ as 
$|a-a_{PM-FM,q=2}| \to 0$.  By the $a \to 1/a$ symmetry, the same is
true of the PM-AFM transition.  The situation at $z=-1$ is more complicated
because the partition function has an isolated zero of multiplicity scaling
like the lattice size there; consequently, as discussed above, the density $g$
has a delta function term $\propto \delta(s)$ as well as its usual term 
(\ref{density}), where $s$ denotes the arclength on ${\cal B}$ away
from the point $z=-1$.  The analysis of Ref. \cite{chitri} found that at
$z=-1$, aside from the leading singularity $\sim (1+z)^{-2}$ in the specific
heat, there is also a subleading logarithmic divergence; it follows that the
density of zeros on ${\cal B}$ near to $z=-1$ has, in addition to the delta
function term, a term that vanishes like $s$.  The zeros in
Figs. \ref{hex2}-\ref{hex2special} are consistent with this. 

\subsection{$q=3$ Case} 

\begin{figure}
\centering
\ 
\epsfxsize=12 cm
\epsfbox{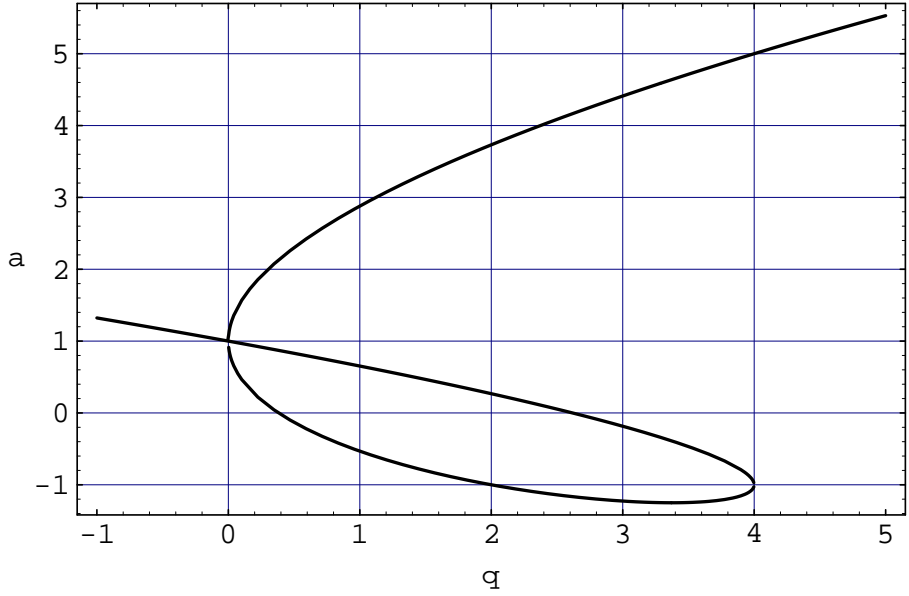}
\caption{Real roots of eq. (\ref{hceqa}), as a function of $q$.}
\label{aRootsHex}
\end{figure}

   For general $q$, from duality and a star--triangle relation, an equation
yielding the value of the PM-FM transition point has been derived \cite{kj},
viz., $x^3-3x-\sqrt{q}=0$, or, in terms of $a$, 
\beq
a^3-3a^2-3(q-1)a-q^2+3q-1 =0
\label{hceqa}
\eeq
For $0 < q < 4$, this equation has three real roots, while for $q > 4$ (and
the formal values $q \le 0$) it has one real root.  The motion of the real
roots as a function of $q$ is plotted in Fig. \ref{aRootsHex}. 
For $q=3$, the solutions are 
\beq
a_{1,q=3} = a_{PM-FM,q=3} = 1 + 2\sqrt{3}\cos(\pi/18) = 4.41147...
\label{ahc1q3}
\eeq
\beq
a_{2,q=3} = 1 - \sqrt{3}\cos(\pi/18) + 3\sin(\pi/18) = -0.1847925...
\label{ahc2q3}
\eeq
and 
\beq
a_{3,q=3} = 1 -\sqrt{3}\cos(\pi/18) - 3\sin(\pi/18) = -1.22668...
\label{ahc3q3}
\eeq
The point $a_{PM-FM,q=3}$ is the physical PM--FM critical point.  
As discussed in
Ref. \cite{p3afhc}, if one follows the roots of eq. (\ref{hceqa}) as $q$ is 
changed continuously, one sees that the middle root $a_2$ decreases from the 
PM--AFM critical point $2-\sqrt{3}$ for $q=2$ through 0 at 
$q_z=(3+\sqrt{5})/2=2.618...$ to the negative value (\ref{ahc2q3}) for $q=3$.
This reflects the fact that as $q$ increases from 2 to $q_z$, the physical 
AFM phase is squeezed out. 

\begin{figure}
\centering
\ 
\epsfxsize=8.5 cm
\epsfbox{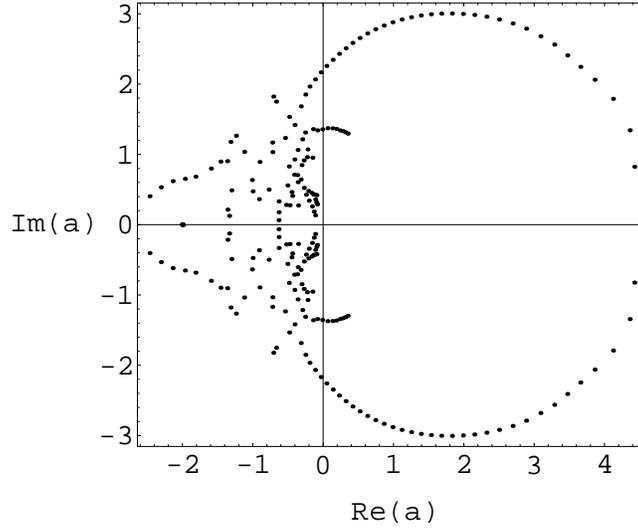}
\caption{Partition function zeros in the $a$ plane for the $q=3$
Potts model on a honeycomb lattice of size $8 \times 6$ hexagons and boundary
conditions of type (fbc,pbc).}
\label{hex3fpa}
\end{figure}

\begin{figure}
\centering
\ 
\epsfxsize=8.5 cm
\epsfbox{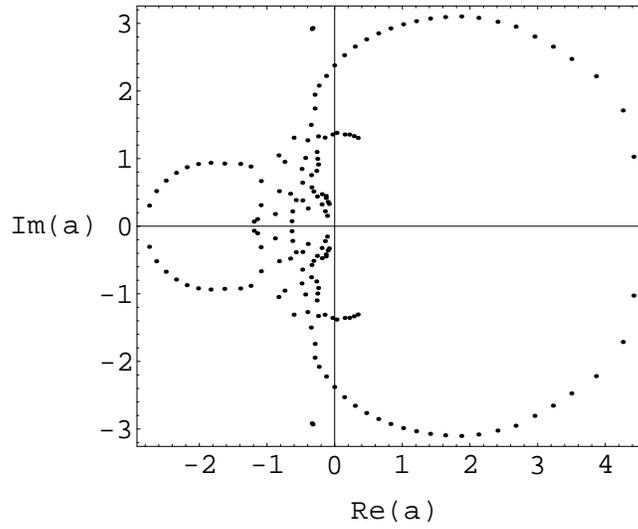}
\caption{Partition function zeros in the $a$ plane for the $q=3$ 
Potts model on a honeycomb lattice of size $8 \times 6$ hexagons and boundary
conditions of type (pbc,fbc).}
\label{hex3pfa}
\end{figure}

\begin{figure}
\centering
\ 
\epsfxsize=8.5 cm
\epsfbox{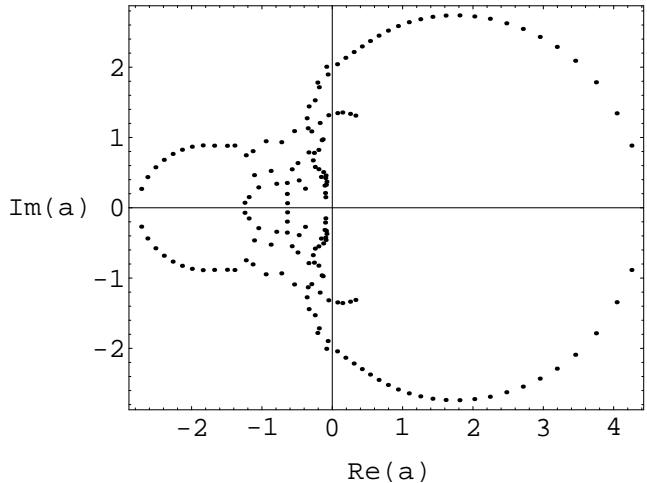}
\caption{Partition function zeros in the $a$ plane for the $q=3$
Potts model on a honeycomb lattice of size $8 \times 6$ hexagons and boundary
conditions of type (pbc,fbc)$_\Delta$.}
\label{hex3pfSpecial}
\end{figure}

Our zeros of the partition function for the $q=3$ case on the honeycomb
lattice are shown in Figs. \ref{hex3fpa} - \ref{hex3pfSpecial} for the three 
types of boundary conditions discussed before. 

In all three plots, the roots $a_{PM-FM,q=3}$ and $a_{3,q=3}$ correspond very 
well to points where the CT phase boundaries comprising ${\cal B}$ 
(as inferred for the thermodynamic limit from
these zeros on finite lattices) cross the real axis.   Hence, 
one anticipates that a CT phase boundary might cross the real axis at the 
value of the middle root, $a_{2,q=3}=-0.1848...$. From the zeros in 
Figs. \ref{hex3fpa} - \ref{hex3pfSpecial},
one can indeed infer that in the thermodynamic limit a CT phase boundary 
curve could cross the real axis at this point.  Since the specific heat
exponent has the known value $\alpha=1/3$ for this model \cite{wurev}, it
follows from eq. (\ref{density}) that the density of zeros near the physical 
PM-FM transition point vanishes like $g(s) \sim s^{2/3}$, where $s$ is the
arclength on ${\cal B}$ away from this point.  This is consistent with the
calculated zeros. 

Further, we see arcs protruding into the PM phase, ending at complex 
conjugate (c.c.) points $a_e, a_e^* = 0.37(2) \pm 1.29(3)i$, where the 
numbers in parentheses refer to the estimated uncertainties in the final 
digits.  Evidently, these
are the analogues for $q=3$ of the exactly known arcs for the Ising $q=2$ 
case.  While the arcs in the Ising case have endpoints on the unit circle at 
angles $\theta=\pm \pi/3$, the endpoints in the $q=3$ case lie slightly farther
out from the origin, at $|a_e| \simeq 1.3$, and have slightly larger angles 
$\theta \sim \pm 75^\circ$.  We find that this trend is true for larger $q$
values also, i.e., $|a_e|$ and $\arg(a_e)$ increase with increasing $q$. 

   In addition, there are at least two more points at which curves of zeros 
cross the real $a$ axis, at $a = a_\ell = -2.77(3)$ 
and at $a = -0.65(2)$.   In Ref. \cite{hcl} it was shown that
if the $q$-state Potts antiferromagnet on the dual lattice $\Lambda_d$ has a
PM-AFM transition at $a_{PM-AFM,\Lambda_d}$, then the dual image of this,
namely, ${\cal D}(a_{PM-AFM,\Lambda_d}) = a_{\ell,\Lambda}$, 
is the leftmost point in the $a$ plane where ${\cal B}$ crosses
the real axis.  Since the PM-AFM point satisfies $ 0 \le a_{PM-AFM,\Lambda_d} 
< 1$, it follows that the dual image $a_{\ell,\Lambda}$ satisfies 
$-\infty < a_{\ell,\Lambda} \le -(q-1)$.  In particular, for $q=3$, this
connection was used, in conjunction with a precisely
measured value of $a_{PM-AFM,t,q=3}$ 
on the triangular ($t$) lattice \cite{adler}
to infer the value of $a_\ell$ for the model on the honeycomb (hc) lattice:
$a_{hc,\ell,q=3}={\cal D}(a_{PM-AFM,t,q=3})=-(2.76454 \pm 0.00015)$, 
i.e., $z_{hc,\ell,q=3}=a_{hc,\ell,q=3}^{-1}=-(0.36172 \pm 0.00002)$.
Our zeros are in agreement with this result.  This point also manifests
itself as a singularity evident from low-temperature series for the specific
heat $C$, magnetization $m$, and susceptibility $\chi$, which yield the value 
$z=-0.363 \pm 0.003$ \cite{jge}.  Using duality and the weakly first order
nature of the physical PM-AFM transition of the $q=3$ Potts antiferromagnet, it
follows that the free energy also has the same weakly first order singularity
at $a_{hc,\ell,q=3}$.  The low-temperature series analysis of Ref. \cite{jge} 
found evidence for a continuous transition at this point, with exponents 
$\alpha_\ell=0.5$, $\beta_\ell=0.11$, and $\gamma_\ell=1.15$.  We have repeated
the series analysis with dlog Pad\'e approximants (PA's) and differential 
approximants (DA's) \cite{tonyg}.  Our DA results also yield 
$\alpha \simeq 0.5$; our PA's did not locate the singularity with sufficient 
precision to infer a reliable value for $\alpha$.
Given the duality and the fact that $\alpha=\alpha'$ for the physical 
PM-AFM transition of the $q=3$ Potts AF
on the triangular lattice, it follows that the singularity in the free energy
of the $q=3$ Potts model on the honeycomb lattice at $a_\ell$ must also be the
same as approached from the right or left.    Since the singularity in 
the internal energy at a singular point 
$a_s$ is $U_{sing} \sim |a-a_s|^{1-\alpha}$, one normally assigns the formal
value $\alpha=1$ to a first-order transition.  A possible way of reconciling
these results is to observe that if a first order 
transition occurs superimposed with a divergent specific heat, then one could
get a value of $\alpha < 1$ in fitting the transition.  For example, 
consider an illustrative internal energy function that behaves near a phase
transition point like 
\beq
U(T) \sim U_{analytic} + c_{1,+}\Theta(T-T_c) + c_{2,+} |T-T_c|^{1/2} 
\label{utest}
\eeq
for $T \searrow T_c$, and similarly for $T \nearrow T_c$, with the coefficients
replaced by $c_{1,-}$ and $c_{2,-}$.  Here, $U_{analytic}$ denotes terms that
are analytic near $T_c$ and $\Theta(x)$ is the step function, 
$\Theta(x)=1$ if $x > 0$ and 0 otherwise.  As one approached $T_c$ from above 
(below) a high-temperature (low-temperature) series analysis would give 
$\alpha=1/2$, but the transition would still be first order because of the 
discontinuous term.  A one-sided version of this behavior occurs in
the in the six-vertex model for the ferroelectric compound potassium 
dihydrogen phosphate (KDP) \cite{kdp}; in that case, the form (\ref{utest}) 
applies for the
high-temperature side, while $U$ is a constant on the low-temperature side.

   Another source of information on $\alpha$ is the density of 
zeros.  However, it is difficult to use this to obtain an accurate
value of $\alpha$.  For example,  Ref. \cite{pfef} included calculations of 
zeros for the $q$-state square-lattice Potts model not just for the values
$q=3,4$ where the PM-FM transition is continuous, but also for the values 
$q=5$ and 6, where this PM-FM transition is first order; see Figs. 3 and 4
therein.  For these cases, one would formally set $\alpha=1$ as mentioned
above, so that eq. (\ref{density}) would predict that the density $g$ of zeros
should remain essentially constant up to the endpoint of the distribution (of
course, the positivity of the coefficients of the terms in the partition
function means that for a finite lattice, there cannot be any zeros on the 
positive real axis in the $a$ or $z$ plane).  
This is consistent with the plots of zeros for these $q=5$
and 6 cases, but it would be difficult to extract an accurate estimate of 
$\alpha$ from those plots.  Below we shall present a similar
plot for another case where the PM-FM transition is known to be first order,
namely the $q=5$ Potts model on the honeycomb lattice, and a similar comment
applies to this plot. 

   Further CT singularities and their relation with the boundary ${\cal B}$
will be discussed elsewhere in work with the authors of Ref. \cite{jge}. 
Our calculations also suggest that there are several unphysical 
O phases which overlap with parts of the negative real $a$ axis. 
There may be other O phases that do not touch the real axis, but the
resolution is not high enough to make a definitive statement here.  

   Concerning the sensitivity of the zeros to lattice boundary 
conditions, several remarks are in order.  The zeros in the $Re(a) \ge 0$ 
half plane are relatively insensitive to these boundary conditions.  However,
certain features of the zeros in the $Re(a) < 0$ half plane do show such 
sensitivity.  This is similar to what was found from a comparative study of
different boundary conditions for the zeros of $Z$ for the $q$-state Potts 
model on the square lattice for several values of $q$ \cite{martinbook,pfef} 
(see also Ref. \cite{martinbc}). 

We have carried out similar calculations of zeros for the $q=4$ Potts
model on the honeycomb lattice, and these will be reported in joint work with 
the authors of Ref. \cite{jge}. 

\subsection{$q=5$ Case}

\begin{figure}
\centering
\ 
\epsfxsize=8.5 cm
\epsfbox{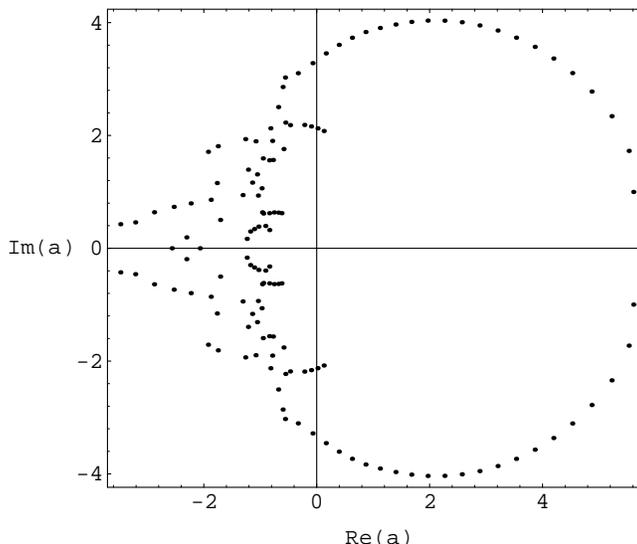}
\caption{Zeros of $Z$ in the $a$ plane for the $q=5$
Potts model on a honeycomb lattice of size $7 \times 6$ hexagons and boundary
conditions of type (fbc,pbc).} 
\label{hex5}
\end{figure}

   It is also of interest to investigate a value of $q$ in the range where the
PM--FM transition is first order, i.e., $q \ge 5$.  We have done this for the
value $q=5$, and we show a resulting plot of zeros in Fig. \ref{hex5}.  
Here, eq. (\ref{hceqa}) has the single real root, which is the PM-FM critical
point, 
\beq
a_{PM-FM,q=5}=2^{-1/3} \cdot 5^{1/2}(1+5^{1/2})^{1/3} + 
2^{1/3} \cdot 5^{1/2}(1+5^{1/2})^{-1/3} + 1 = 5.5298...
\label{ahc1q5}
\eeq
Since the Potts antiferromagnet with $q=3$ and $q=4$ on the triangular lattice
has, respectively, a finite-temperature PM-AFM phase transition
\cite{grest,adler} and a zero-temperature critical point \cite{baxter87}, 
it is expected that for $q \ge 5$, the model is disordered for all
temperatures.  This, together with the connection discussed in Ref. \cite{hcl},
would imply that the leftmost point at which ${\cal B}$ crosses the real axis 
for the $q=5$ Potts model on the honeycomb lattice is $a_\ell < -4$.  Our zeros
are consistent with this. 

\subsection{Further Discussion} 

   It is a general feature of the maximal (or sole) real solution of
eq. (\ref{hceqa}), i.e., $a_{PM-FM}$, 
that it increases monotonically with $q$ for $q \ge 0$.  This
is evident in Fig. \ref{aRootsHex} and reflects the basic thermodynamic
property that as $q$ increases, the spins become ``floppier'', and one must go
to lower temperature to obtain FM long range order. 
In addition to the features already discussed, we note that (i) the leftmost
point where ${\cal B}$ crosses the real $a$ axis, $a_\ell$, moves to the
left as $q$ increases; and (ii) the points where ${\cal B}$ crosses the
imaginary axis move out from the origin as $q$ increases.  Both of these
features can be understood, as discussed directly above, by the reduction in
the size of the (FM) ordered phase as $q$ increases. 

   Because of the duality relation, these partition function zeros, in the
$a$ plane, of the $q$-state Potts model on the honeycomb lattice also yield 
equivalent zeros of the same model on the dual, triangular lattice in the
plane of the variable $a_d$ given in eq. (\ref{ad}).  A comparison of the 
plots calculated with different boundary conditions is valuable since this
gives a measure of the effects of these boundary conditions (see also Refs. 
\cite{mm,martinbc}).  

\section{Partition Function Zeros on the Kagom\'e Lattice}

\subsection{Comparison with Exact ${\cal B}$ for Ising $q=2$ Case}

\begin{figure}
\centering
\
\epsfxsize=9.5 cm
\epsfbox{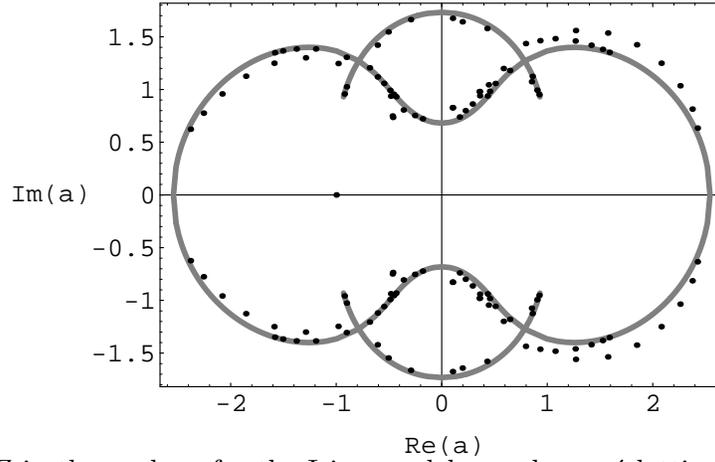}
\caption{Zeros of $Z$ in the $a$ plane for the Ising model on a kagom\'e 
lattice with $4 \times 6$ hexagons and (pbc,fbc) boundary conditions.}
\label{kag2plain}
\end{figure}
\begin{figure}
\centering
\
\epsfxsize=9.5 cm
\epsfbox{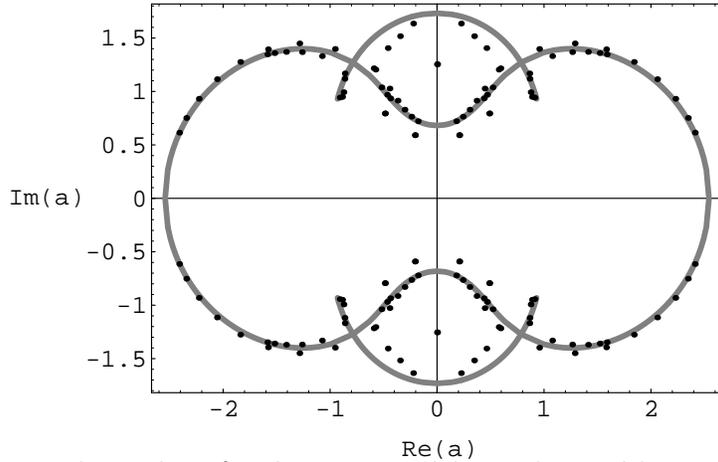}
\caption{Zeros of $Z$ in the $a$ plane for the Ising model on a kagom\'e
lattice with $4 \times 6$ hexagons and (pbc,fbc)$_\Delta$ boundary 
conditions.}
\label{kag2special}
\end{figure}

   For our calculations of zeros of Potts model on the kagom\'e lattice, 
two of the boundary conditions that we use are (pbc,fbc) and
(pbc,fbc)$_\Delta$, where now $\Delta=4$.  The third can be described as
follows: we start on the lattice that is the dual to kagom\'e, namely the
diced lattice, and impose (pbc,fbc) boundary conditions; then we
transform the results back to the kagom\'e lattice by the duality map on $a$,
eq. (\ref{ad}).  To save space, for each value of $q$, 
we only show results for the first two of these choices of boundary 
conditions.  For $q=2$ these are given in Figs. 
\ref{kag2plain} and \ref{kag2special}. The exact CT phase boundary ${\cal B}$
is given by the locus of solutions of the equation 
\beq
a^8 + 18a^4 + 24a^2 + 21- 4(1+a^2)(1-a^2)^2p=0 
\label{kagisingboundary}
\eeq
where $-3/2 \le p \le 3$ \cite{pnote}.  
Because the coordination number of the kagom\'e
lattice is even, this locus is symmetric under $a \to -a$.  
In Ref. \cite{cmo}, the locus was plotted in the $z$ and $u=z^2$ 
planes (see also Ref. \cite{dotera}).  Here it is shown as the gray curves in
the $a$ plane, consisting of a ``dumbell'' part and a complex conjugate pair of
circular arcs which intersect the dumbell at four multiple points 
(the analytic expressions for
which are given in Ref. \cite{cmo}).  The inside of the dumbell region is the
PM phase, the c. c. regions between the narrow neck of the dumbell and the
circular arcs are O phases, and the region outside of ${\cal B}$ and extending
to complex infinity is the (CT extension of the) FM phase.  The PM-FM critical
point is given by 
$a_{kag,PM-FM,q=2} = -a_{kag,\ell,q=2} = 3^{1/4}(2-\sqrt{3})^{-1/2}=2.542...$ 
Just there is no physical AFM phase (owing to the frustration of the Ising
AF on the kagom\'e lattice), so also there is no complex-temperature extension
thereof.  Comparing the zeros calculated with the
different boundary conditions, we find that with the (pbc,fbc) choice, the
zeros on the neck of the dumbell and on the outer circular arcs track the exact
curves well, while those on the right (left) lie slightly outside (inside) the
CT phase boundaries.  With the (pbc,fbc) boundary conditions, not all sites
have even coordination number, so that $Z$ contains some odd powers of $a$, 
and hence the $a \to -a$ symmetry of the exact boundary is not precisely 
maintained by the zeros.  In passing, we note that because the sites on the
upper and lower boundaries have odd coordination number $\Delta=3$, theorem 
6 of Ref. \cite{cmo} implies that $Z(z=-1)=0$, and this zero (which is
multiple) is evident in Fig. \ref{kag2plain}.  
For the (pbc,fbc)$_\Delta$ boundary conditions, (i) most of 
the zeros near to the dumbell lie closer to the exact curves, but the zeros
near the arcs lie farther away from them, as compared with the situation for
the (pbc,fbc) choice; and (ii) $Z$, and hence its zeros, is invariant under the
negation $a \to -a$, in contrast to case with the (pbc,fbc) case; (iii) because
all sites have even coordination number,  there is no zero in $Z$ at $z=-1$. 
For both types of boundary conditions, the density of zeros in the vicinity of
the PM-FM critical point $a_{kag,PM-FM,q=2}$ decreases in a manner consistent 
with the form from eq. (\ref{density}) with $\alpha=0$ for the 2D Ising model,
viz., $g \sim s$ as $s \to 0$, where $s$ the arclength along ${\cal B}$ 
away from $a_{PM-FM,q=2}$.

\subsection{$q=3$ Case}

\begin{figure}
\centering
\ 
\epsfxsize=8.5 cm
\epsfbox{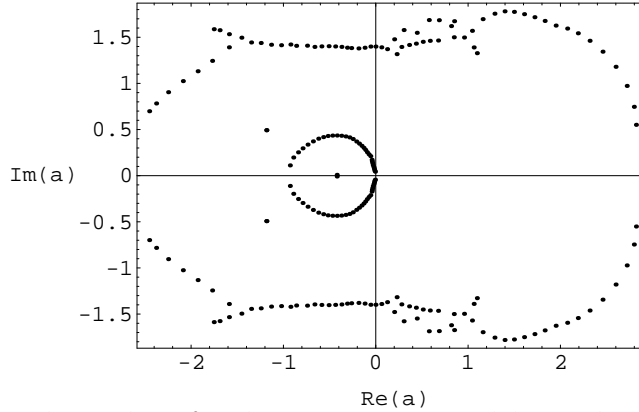}
\caption{Zeros of $Z$ in the $a$ plane for the $q=3$ Potts model 
on a kagom\'e lattice of size $4 \times 8$ hexagons and (pbc,fbc) boundary
conditions.}
\label{kag3plainA}
\end{figure}
\begin{figure}
\centering
\ 
\epsfxsize=8.5 cm
\epsfbox{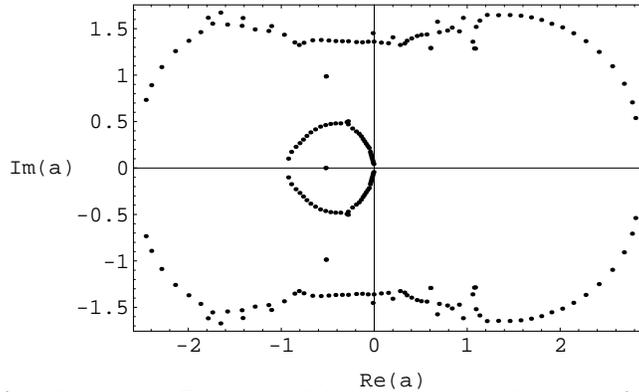}
\caption{Zeros of $Z$ for the $q=3$ Potts model on a kagom\'e lattice of 
size $4 \times 8$ hexagons and (pbc,fbc)$_\Delta$ boundary conditions.}
\label{kag3specialA}
\end{figure}

   We show our zeros of $Z$ for the $q=3$ Potts model on the kagom\'e
lattice in Figs. \ref{kag3plainA} and \ref {kag3specialA}.
In this case we use lattices of sizes $N_x \times N_y$ with $N_y$ larger 
than $N_x$ in order to compensate for the fact that the free boundaries are in
the $y$ direction and free, as contrasted with periodic, boundary conditions
introduce greater finite-size effects. 
The zeros suggest that in the thermodynamic limit, the inferred CT phase 
diagram for the $q=3$ kagom\'e lattice may involve somewhat simpler boundary
curves than was the case for the same model on the honeycomb lattice. 
There is a high--temperature PM phase, a low--temperature FM phase, and there 
are strong
indications of a third CT phase whose right--hand boundary crosses the real
axis at $a=0$, corresponding to a zero--temperature critical point of the $q=3$
Potts antiferromagnet on this lattice.  This is in good agreement with the
known property that this model does have such a $T=0$ critical point (which can
be related to the $T=0$ critical point of the $q=4$ Potts antiferromagnet on
the triangular lattice) \cite{kag3t0,henley}.  The 
inferred position where the CT boundary crosses the real axis on the right is
at $a_{kag,PM-FM,q=3}=2.84(4)$. This is in accord, to within the 
uncertainty, with the value of $a_{kag,PM-FM,q=3}=2.87646(4)$, i.e., 
$z_{kag,PM-FM,q=3}=0.347650(5)$) obtained from series analysis \cite{jge}.
The left--hand boundary of the third phase crosses the real axis at about 
$a=-0.96(3)$.  From Fig. \ref{kag3specialA}, where the boundary between the 
(CT extensions of the) PM and the FM phases is probably best represented, 
we infer that the leftmost point where this CT phase boundary crosses the real 
axis is at $a_{kag,\ell,q=3}=-2.54(6)$.  
This point is manifested as a singularity in
thermodynamic quantities evident in low-temperature series analysis, which
obtains $a_{kag,\ell,q=3}=-2.486(3)$ (i.e., $z_{kag,\ell,q=3}=-0.4023(5)$).  
Although the CT phase boundary is not symmetric under 
$a \to -a$ as was true for $q=2$ on this lattice, one can still discern a 
remnant of the dumbell shape that occurred for the $q=2$ case. 
As before for the honeycomb lattice with $q=3$, the $g \sim s^{2/3}$ 
decrease in the density of zeros in the vicinity of the PM-FM critical point is
consistent with the calculated zeros. 

\subsection{Case $q=4$} 

\begin{figure}
\centering
\ 
\epsfxsize=8.5 cm
\epsfbox{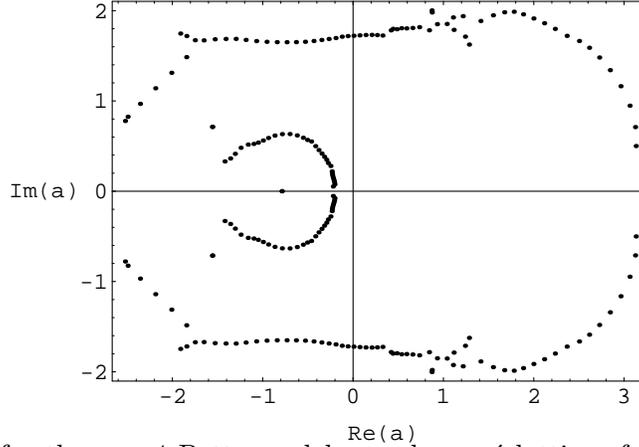}
\caption{Zeros of $Z$ for the $q=4$ Potts model on a kagom\'e lattice of
size $4 \times 8$ hexagons and (pbc,fbc) boundary conditions.}
\label{kag4plainA}
\end{figure}

\begin{figure}
\centering
\ 
\epsfxsize=8.5 cm
\epsfbox{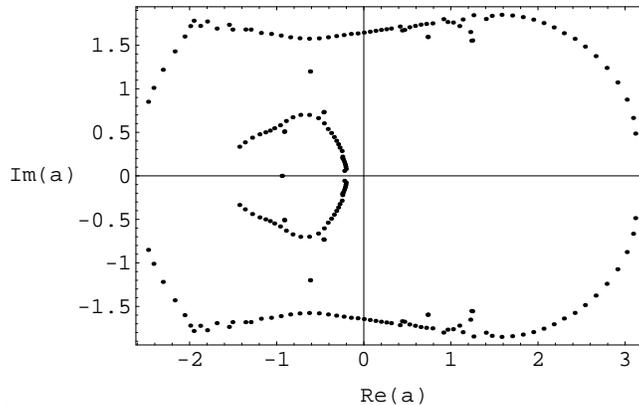}
\caption{Zeros of $Z$ for the $q=4$ Potts model on a kagom\'e lattice of
size $4 \times 8$ hexagons and (pbc,fbc)$_\Delta$ boundary conditions.}
\label{kag4specialA}
\end{figure}

For $q=4$, we present our results in the Figs. \ref{kag4plainA} and 
\ref{kag4specialA}.  The main differences between the locus of zeros, as
compared with the case of $q=3$ is, first, that the crossing which occurs at 
$a=0$ for $q=3$ is shifted to a negative value, $a=-0.21(2)$ for $q=4$. This
demonstrates that the $q=4$ Potts antiferromagnet on the kagom\'e lattice 
has no PM-AFM phase
transition (or any hypothetical non-symmetry breaking phase transition) at 
finite temperature or any critical point at $T=0$.  This conclusion also
follows for $q=5$ since increasing $q$ beyond 4 has the effect of making 
the spins ``floppier'' and the model more disordered.  For $q \ge 6$, this
conclusion has been proved rigorously \cite{sokal}.  Second, it appears that 
the previously presumably closed inner ring of zeros has now opened at 
its leftmost point, which would imply that now there would be only two 
phases (with their CT extensions), the PM and FM.  The values inferred for 
$a_{kag,PM-FM,q=4}$ and $a_{kag,\ell,q=4}$ are in accord with the values
obtained from series analysis \cite{jge}.  Since the
specific heat critical exponent $\alpha=2/3$ for the $q=4$ Potts model on 2D
lattices, eq. (\ref{density}) gives $g \sim s^{1/3}$ for the manner in which
the density of zeros vanishes as one approaches the PM-FM critical point along
the CT phase boundary.  In particular, this implies that the decrease in
density should be less rapid for $q=4$ than for $q=3$, and, indeed, this is
evident from a comparison of our plots of zeros for these two cases on the
kagom\'e lattice. 

\subsection{Partition Function Zeros on the Diced Lattice}

   As with the honeycomb lattice and its dual, the triangular lattice, our 
zeros, in the $a$ plane, of the partition function for the $q$-state Potts 
model on the kagom\'e lattice also yield equivalent zeros of the same 
model on the lattice that is dual to the kagom\'e lattice, in the plane of the
variable $a_d$ given in eq. (\ref{ad}).  Henceforth, we shall suppress the
subscript $d$ on $a_d$.  This dual lattice is called the diced
lattice; as discussed above, it is not an Archimedean lattice \cite{gs}; 
rather, it is a tiling of the plane with identical rhombi such that, as one 
traverses a circuit along the edges of each rhombus, one passes
vertices with coordination number 3,6,3,6 in sequence.  Thus, in standard
mathematical notation, the diced lattice is the lattice $[3 \cdot 6 \cdot 3
\cdot 6]$ dual to the $(3 \cdot 6 \cdot 3 \cdot 6)$ (= kagom\'e) lattice. 
Some relevant properties of the diced lattice are noted in (Table II
of) Ref. \cite{wn}.  Although the faces of the diced lattice are identical, the
vertices are not (this is the dual of the property that the vertices of an
Archimedean lattice are identical but the faces are, in general, not, since
an Archimedean lattice can consist of more than one type of regular polygon).
In particular, the diced lattice has vertices of two different types: one 
with an odd degree (= coordination number) $\Delta=3$, and the other with
even degree, $\Delta=6$.  Indeed, the diced lattice is bipartite, and its two 
sublattices, which we may denote $\Lambda_3$ and $\Lambda_6$, are comprised 
of the vertices with degree $\Delta=3$ and $\Delta=6$, respectively.  The
vertices in the $\Lambda_3$ and $\Lambda_6$ sublattices occupy the respective 
fractions $f_3 = 2/3$ and $f_6=1/3$ of all the vertices.  This is quite 
different from bipartite Archimedean lattices, where the vertices on each
of the two sublattices occupy the same fraction, $f=1/2$, of the total number
of vertices (as a consequence of the fact that on an Archimedean lattice, all
vertices are equivalent).  

\begin{figure}
\centering
\
\epsfxsize=8.5 cm
\epsfbox{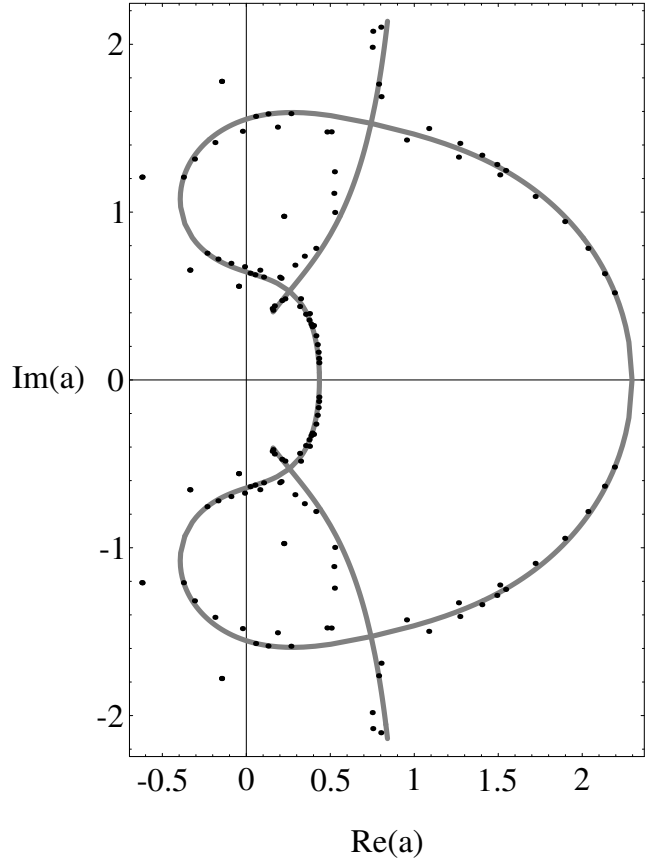}
\vspace{1 cm}
\caption{Zeros of $Z$ for the Ising model on a diced lattice,
obtained via duality from a kagom\'e lattice of size
$4 \times 8$ hexagons and (pbc,fbc)$_\Delta$ boundary conditions.}
\label{diced2special}
\end{figure}

   For our discussion of the CT phase diagrams of the $q$-state Potts model
with $q=3,4$ on the diced lattice, it is instructive to begin by discussing 
the $q=2$ case, for which one can use exact results on the free energy.  The CT
phase diagram is shown in Fig. \ref{diced2special}.  
The exact CT phase boundary ${\cal B}$ \cite{dotera,cmo} is
shown as the dark solid curve.  Using duality and the $z \to -z$ symmetry of 
the boundary for the Ising model on the kagom\'e lattice (the latter of
which follows from the even coordination number of that lattice), it follows
that the ${\cal B}$ in Fig. \ref{diced2special} for the Ising model on the 
diced lattice is the same as ${\cal B}$ in the $v$ plane for the model on the 
kagom\'e lattice, where $v=(1-z)/(1+z)$.  Note that although the 
physical FM and AFM phases are
disjoint, the respective complex-temperature extensions of these phases are 
analytically connected.  The reason for this is that, in contrast to 
bipartite Archimedean lattices, the two sublattices $\Lambda_3$ and
$\Lambda_6$ of the diced lattice do not occupy the same fraction of the total
lattice.  Thus, reverting to conventional Ising model notation for this
discussion, let us define $M_{\Lambda_3}$ and $M_{\Lambda_6}$ as the
magnetizations of the sublattices $\Lambda_3$ and $\Lambda_6$ and $M_{unif.}$
and $M_{stag.}$ as the uniform and staggered magnetizations, all per unit
area of the total lattice, with 
\beq
M_{unif.} = M_{\Lambda_3} + M_{\Lambda_6}
\label{muniform}
\eeq
\beq
M_{stag.} = M_{\Lambda_3} - M_{\Lambda_6}
\label{mstaggered}
\eeq
If the present lattice had been Archimedean, with each sublattice occupying a
fraction 1/2 of the total, then $M_{stag.}$ would vanish identically not just
in the PM phase but also in the FM phase, and $M_{unif.}$ would vanish
identically not just in the PM phase but also the AFM phase, so that the FM and
AFM phases, and their complex-temperature extensions, could not be analytically
connected with each other.  However, because the sublattices of the diced
lattice occupy different fractions of the total lattice, it follows that in the
limit of complete sublattice spin ordering, $M_{\Lambda_3}=2/3$ and 
$M_{\Lambda_6}=1/3$ and hence, besides the obvious result, 
$M_{unif.}(a=\infty)=1$, one has 
\beq
M_{unif.}(a=0)=\frac{1}{3} 
\label{muniforma0}
\eeq
\beq
M_{stag.}(a=\infty)=\frac{1}{3}
\label{mstagainf}
\eeq
\beq
M_{stag.}(a=0)=1
\label{mstaga0}
\eeq
That is, the uniform magnetization $M_{unif.}$ does not vanish even in the
region of complete sublattice magnetizations of opposite sign, at $T=0$ for 
$J < 0$, i.e., $a=0$, and the staggered magnetization $M_{stag.}$ does not 
vanish even in the limit of complete sublattice magnetizations of the same 
sign, at $T=0$ for $J > 0$, i.e., $a=\infty$.  Hence, there exist paths that
connect the points $a=\infty$ and $a=0$ in the complex $a$ plane.  Of course,
if one restricts to the physical temperature interval $0 \le a \le \infty$,
then the physical FM and AFM phases cannot be analytically connected, since
they are separated by the PM phase, where both $M_{unif.}$ and $M_{stag.}$
vanish identically.  However, the complex-temperature extensions of the FM 
and AFM phases are analytically connected, as is shown by the existence of the
paths alluded to above. 

    The CT phase diagram is thus as follows: (see Fig. 
\ref{diced2special}: first, 
there is a symmetric, high-temperature PM phase around 
the point $a=1$ that includes the interval $a_1 < a < a_1^{-1}$ on the real
axis, where 
\beq
a_1 = \frac{1}{2}(1+\sqrt{3})\Bigl [1-(2\sqrt{3}-3)^{1/2}\Bigr ]=0.43542...
\label{a1}
\eeq
Secondly, there is the single complex-temperature extension of the two 
different physical FM and AFM phases; this extension includes the intervals 
$-\infty \le a < a_1$ and $a_{1}^{-1} < a \le \infty$ on the real axis
(see eq. (4.12) in Ref. \cite{cmo}) and extends outward to complex infinity in
the $a$ plane.   We label this phase as CT(A)FM. 
Third, there is a complex-conjugate pair of O phases.  In
Fig. \ref{diced2special} we have shown the zeros computed with one
particular set of boundary conditions; in this case and also with the other
types of boundary conditions, these zeros agree well with the exact 
results.  (This is dual to the same statement for the $q=2$ kagom\'e lattice.) 

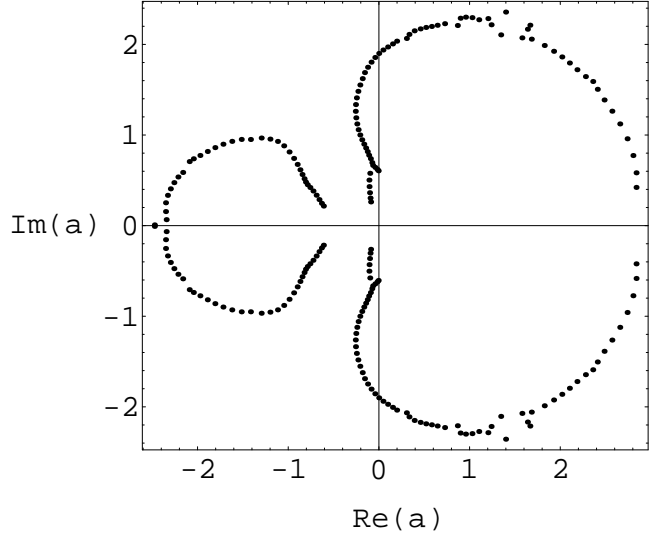
\begin{figure}
\centering
\
\epsfxsize=8.5 cm
\epsfbox{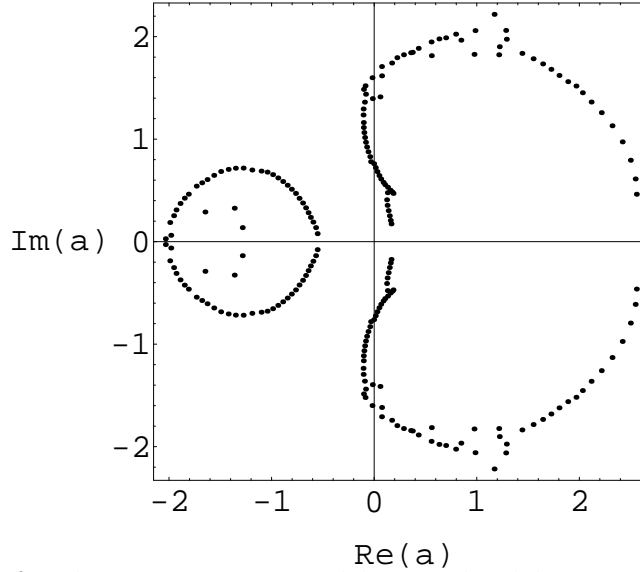}
\caption{Zeros of $Z$ for the $q=3$ Potts model on a diced lattice with 
(pbc,fbc) boundary conditions and of size equivalent to $4 \times 8$ hexagons 
on a kagom\'e lattice.} 
\label{diced3dual}
\end{figure}

    Proceeding to the cases that have not been exactly solved, in  Fig. 
\ref{diced3dual} we show our zeros for the $q=3$ Potts model on the
diced lattice, obtained via duality from those on the kagom\'e lattice.  For
this and $q=4$, we show results with only one set of boundary conditions, since
the other boundary conditions yield similar results.  As noted in
Ref. \cite{hcl}, from the finding in Ref. \cite{jge} of a CT singularity at 
$z_\ell=-0.4023(5)$ in the $q=3$ Potts model on the kagom\'e lattice, it
follows, using the duality connection, that the $q=3$ Potts model has a phase
transition from the PM phase to the FM-AFM phase at the point 
$a_{diced,q=3,PM-AFM}=0.1393(8)$.  This constitutes the left border of the
physical PM phase on the positive real $a$ axis. 
Moreover, again by duality, from the PM-FM
transition point of the model on the kagom\'e lattice, determined from series
analysis in Ref. \cite{jge} to be at $z_c=0.347650(5)$, it follows that the
position of the PM-FM transition of the $q=3$ Potts model on the diced lattice
is at $a_{diced,PM-FM,q=3}=2.59876(4)$.  As in the $q=2$ case, and for the same
reason, although the physical FM and AFM phases are disjunct, their 
complex-temperature extensions are analytically connected.  We thus again label
this extension as the CT(A)FM phase. 
The other CT phases include the extension of 
the PM phase and an O phase in the $Re(a) < 0$ half-plane.  Our finding that
the CT phase boundary for the $q=3$ Potts model on the kagom\'e lattice has a
component that passes through $a=0$, corresponding to a zero-temperature
critical point in that model, implies, by duality, that the boundary of the O
phase in the model on the diced lattice crosses the real $a$ axis on the left
at $a=-2$. 

   We show our zeros for the $q=4$ Potts model on the diced lattice in Fig. 
\ref{diced4dual}.  For this case, from the value 
$a_{kag,PM-FM,q=4}=3.1561(5)$ obtained from series analysis in Ref. 
\cite{jge}, we deduce, using duality, that the PM-FM critical point for the
diced lattice is $a_{diced,PM-FM,q=4}=2.8552(5)$.  Further, from the value
obtained for the CT singularity, $z_{kag,q=4,\ell}=-0.42 \pm 0.01$ \cite{jge},
we have deduced, again using duality, that the $q=4$ Potts antiferromagnet on 
the diced lattice has no finite-$T$ phase transition and also is not critical 
at $T=0$, since 
\beq
{\cal D}(a_{kag,q=4,\ell})=-(0.18 \pm 0.02)
\label{adicedq4}
\eeq
is negative \cite{hcl}.  There is thus no AFM phase for $q=4$. Because
increasing the value of $q$ for a fixed temperature makes the spins floppier,
this result implies that there is also no AFM phase for $q \ge 5$.  
In the context of the complex-temperature phase diagram, the point in eq. 
(\ref{adicedq4}) corresponds to the point where the two arcs close in the
thermodynamic limit and the left-hand boundary of the PM phase crosses 
the real $a$ axis in Fig. \ref{diced4dual}.  In this figure one also sees a
curve in ${\cal B}$ in the $Re(a) < 0$ half-plane.  

\begin{figure}
\centering
\
\epsfxsize=8.5 cm
\epsfbox{diced4dual.epsi}
\caption{Zeros of $Z$ for the $q=4$ Potts model on a diced lattice with
(pbc,fbc) boundary conditions and of size equivalent to $4 \times 8$ hexagons
on a kagom\'e lattice.}
\label{diced4dual}
\end{figure}

\section{Conclusions}

   We have calculated complex-temperature zeros of the partition function 
for the $q$-state Potts model on the honeycomb and kagom\'e lattices.
These results give useful information about the complex--temperature
phase diagrams and singularities of these models.  

\vspace{2mm}

\begin{center}
{\bf Acknowledgments} 
\end{center}

   This research was supported in part by the NSF grant PHY-97-9722101. R.S. 
thanks. Prof. A. J. Guttmann for kindly giving us a copy of Ref. \cite{jge} 
prior to publication and for discussions of that work.


\begin{thebibliography}{99}
\bibitem{com}{The free energy also has an isolated singularity at
$|K|=\infty$ and (see Theorem 6 of Ref. \cite{cmo}) for the Ising model on
lattices with odd coordination number, a singularity at $z=-1$.  The latter
lies on the complex-temperature phase boundary for the honeycomb lattice but is
isolated for the heteropolygonal $3 \cdot 12^2$ lattice.}
\bibitem{cte}{As discussed in Ref. \cite{chisq}, the complex-temperature
extension of a physical phase is obtained by analytically continuing the free 
energy from the interval of physical temperature to a maximal region allowed by
nonanalytic boundaries.  Henceforth, we shall generally take the adjective 
``complex-temperature extension'' as implicit when referring to phases.  There
are also other complex-temperatures that have no overlap with any physical
phase; we shall denote these by O for ``other''.}
\bibitem{aap}{For the physical critical point, one does not have to 
distinguish between the exponents $\alpha$ and $\alpha'$ describing
the approach from the high-temperature and low-temperature sides, since these 
are equal.}
\bibitem{pnote}{In both the honeycomb and kagom\'e cases, the respective
conditions (\ref{hcisingboundary}) and (\ref{kagisingboundary}) are the
conditions that $L(a,p)=0$, where $L$ occurs in the term 
$\propto \int_{0}^{2\pi}d\theta_1\int_{0}^{2\pi}d\theta_2 
\ln[L(a,p)]$ in the free energy, and 
$p=\cos(\theta_1)+\cos(\theta_2)+\cos(\theta_1+\theta_2)$.
The range of $p$ is then $-3/2 \le p \le 3$, as given.}
\bibitem{anote}{By duality, for the Ising model on the honeycomb lattice, 
${\cal B}$ in $a$ plane is the same as ${\cal B}$ in the 
$a_d = (a+1)/(a-1)$ plane for same model on the triangular lattice 
\cite{chitri}.  Note that for the case of anisotropic spin-spin couplings, 
the partition function zeros fill out regions rather than lying on curves in 
the $a$ plane \cite{aniso}.}
\end{thebibliography}
\end{document}